\documentclass[preprint,amsmath,amssymb]{revtex4}

\usepackage{txfonts}    
\usepackage{color}     
\usepackage{graphicx}   
\usepackage{bm}

\newcommand{\sub}[1]{\hspace{-.07em}\bm{#1}}

\newcommand{\mtr}[1]{\mathrm{#1} }

\newcommand{\pdif}[2]{\ensuremath{\frac{\partial #1}{\partial #2}}}

\newcommand{\D}{\cdot}

\newcommand{\lbr}{\left ( }
\newcommand{\rbr}{\right ) }

\newcommand{\vpr}{\varv_{\parallel} }

\newcommand{\bb }{\bm{b} }

\newcommand{\vti }{\varv_{\mathrm{ti}} }

\newcommand{\dfg }{\delta f^{\mathrm{(g)}} } 
\newcommand{\dfgk }{\dfg_{\mtr{s}\sub{k}_{\perp}} }

\newcommand{\dphi }{\delta \phi }
\newcommand{\dphik }{\delta \phi_{\sub{k}_{\perp}} }
\newcommand{\dpsi }{\delta \psi }
\newcommand{\dpsik }{\delta \psi_{\sub{k}_{\perp} }}
\newcommand{\dalk }{\delta A_{\parallel \sub{k}_{\perp}} }

\begin{document}


\title{Impact of geodesic curvature on zonal flow generation \\in magnetically confined plasmas}
%
\author{Motoki Nakata$^\#$}
\affiliation{National Institute for Fusion Science, Toki 509-5292, Japan / \\
The Graduate University for Advanced Studies, Toki, 509-5292, Japan, \\
PRESTO, Japan Science and Technology Agency, Saitama 332-0012, Japan}
\thanks{$\#\,$Present affiliation: Faculty of Arts and Sciences, Komazawa University, Tokyo 154-8525, Japan \\ \textbf{This is the author accepted manuscript (AAM/AM), published in Plasma Fusion Research 17, 1203077 (2022)\ \  doi: 10.1585/pfr.17.1203077 \\
\ (with permission from JSPF)}} 
\author{Seikichi Matsuoka}
\affiliation{National Institute for Fusion Science, Toki 509-5292, Japan / \\
The Graduate University for Advanced Studies, Toki 509-5292, Japan}
%
%
%

\begin{abstract}
The impact of magnetic geometry on zonal-flow generation in ion temperature gradient driven turbulence is investigated by means of 
linear and nonlinear gyrokinetic simulations. 
The modulation of geodesic curvature on various configurations has revealed 
amplification of the zonal-flow intensity in relatively smaller geodesic curvature. 
Based on these findings, a nonlinear proxy model for explorations of novel magnetic geometry to activate the zonal-flow dynamics is proposed. 
\end{abstract}

\maketitle
%
Magnetically confined plasma is regarded not only as the medium of fusion reactions, but also as an attractive non-equilibrium 
system with an extreme spatial inhomogeneity/gradient embedded in magnetic fields with the geometric flexibilities. 
One of the most prominent physical processes is spontaneous emergence of mesoscopic coherent structures such as zonal flows, which are nonlinearly 
excited in microscopic turbulence. 
The zonal-flow generation is now widely recognized as a key mechanism leading to improved plasma confinement. 
An important earlier finding by gyrokinetic simulations is the suppression of turbulent transport by enhanced zonal flows\cite{thw_zf,nakata_prl}. 

By fully utilizing the diversity of the three-dimensional magnetic field structures, 
both axisymmetric and non-axisymmetric plasmas have much capabilities to suppress/optimize the turbulence. 
In several pioneering earlier works, turbulence optimization scheme of ``proxy approach'' has been developed, 
where the proxy model $\gamma_{k_\perp}^{\mtr{proxy}}/k_{\perp}^2$ is constructed as a function of the geometric quantities to qualitatively reproduce 
the linear mixing-length diffusivity for the ion temperature gradient(ITG) driven mode and the trapped electron mode(TEM)\cite{prox1,prox2,prox3}. 

In this study, a key geometric dependence to amplify the zonal-flow intensity is extracted by means of the linear 
and nonlinear gyrokinetic Vlasov simulations of the ITG-driven turbulence in non-axisymmetric and axisymmetric plasmas. 
Particular focus is put on the geodesic curvature $\mathcal{K}_{\mtr{g}}$ of the field line on the flux surface, 
where the radial drift excursion for trapped particle is proportional to $\mathcal{K}_{\mtr{g}}$. 
%
Then, the linear damping of zonal flows, which is attributed to the shielding of the zonal potential 
by the radial orbit width of trapped ions, can also affect the zonal-flow intensity in the nonlinear state.  
In addition to the linear zonal-flow response\cite{rh1,rh2,rh3,rh4}, the impact of the geodesic curvature 
on the nonlinearly generated zonal flows are examined here. 
%
%
\begin{figure}[t]
  \centering
  \includegraphics[scale=1.3]{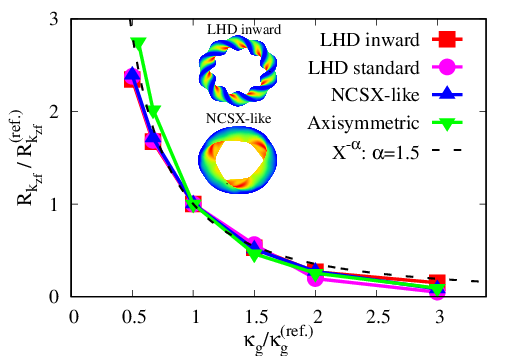}
  \caption{Geodesic curvature dependence of residual ZF level.}
  \label{fig1}
\end{figure}

The governing equation solved in GKV, a multi-species electromagnetic gyrokinetic Vlasov code\cite{gkv}, is briefly shown as follows: 
%
\begin{eqnarray} 
&& \left ( \pdif{}{t} + \vpr \bb \! \D \! \nabla +i\omega_{\mtr{Ds}} - \frac{\mu\bb \! \D \! \nabla B}{m_{\mtr{s}}}\pdif{}{\vpr}
\right ) \delta g_{\mtr{s}\sub{k}_{\perp}}  \nonumber \\
&& -\, \frac{c}{B} \! \sum_{\Delta} \bb \! \D \! \lbr \bm{k}_{\perp}^{\prime} \! \times \! \bm{k}_{\perp}^{\prime \prime} \rbr 
\dpsi_{\mtr{s}\sub{k}_{\perp}^{\prime}} \delta g_{\mtr{s} \sub{k}_{\perp}^{\prime \prime}}   -\, \mathcal{C}_{\mtr{s}} \! \left ( \delta g_{\mtr{s}\sub{k}_{\perp}} \right )  \nonumber \\
&& = \frac{e_{\mtr{s}}F_{\mtr{Ms}}}{T_{\mtr{s}}}\left (
\pdif{\dpsik}{t} 
+ i\omega_{\ast T\mtr{s}}\dpsi_{\mtr{s}\sub{k}_{\perp}}
+\vpr\frac{\mu \bb \cdot \nabla B}{T_{\mtr{s}}}\dpsi_{\mtr{s}\sub{k}_{\perp}}
 \right )\ ,
\end{eqnarray}
%
where $\delta g_{\mtr{s}\sub{k}_{\perp}}=\delta g_{\mtr{s}\sub{k}_{\perp}}(t,z,\vpr,\mu)$ stands for the non-adiabatic part of 
the perturbed gyrocenter distribution function $\dfgk$ for the particle species ``s''.
The gyro-averaged potential fluctuation is denoted by 
$\dpsik \! := \! J_{0\mtr{s}}[\dphik \! - (\vpr/c)\dalk]$, 
where the former and latter terms mean the electrostatic and electromagnetic parts, respectively.
The magnetic drift frequency $\omega_{\mtr{Ds}}$ is expressed by 
%
\begin{eqnarray} 
\omega_{\mtr{Ds}} =   \frac{c}{e_{\mtr{s}}B}\bm{k}_{\perp} \! \D \bb \times 
\lbr \mu \nabla B + m_{\mtr{s}}\vpr^{2}\bb \! \D \! \nabla \bb \rbr \nonumber \\ 
 =  \frac{c ( m_{\mtr{s}}\vpr^2 \! +\! \mu B )}{e_{\mtr{s}}B_{\mtr{ax}}}\lbr \mathcal{K}_{\mtr{g}}k_{x}+\mathcal{K}_{\mtr{n}}k_{y} \rbr\ ,
\end{eqnarray}
%
where $\mathcal{K}_\mtr{g}$ and $\mathcal{K}_\mtr{n}$ represents the geodesic and normal curvature components, respectively.  
More detailed expressions are available in e.g., Refs. 11 and 12. 

The local fluxtube gyrokinetic simulations of the linear zonal-flow response and the ITG-driven turbulence have been carried out 
for the several magnetic configurations, i.e., Large Helical Device(LHD), NCSX-like\cite{ncsx}, and the axisymmetric cases, 
where the adiabatic electron response is assumed. 
%
The radial position of $\rho=0.5$ with the safety factor 
$q(\rho\!=\!0.5) \! = \!$ \{1.99(LHD inward), 2.16(LHD standard), 2.14(NCSX-like), 1.41(Axisymmetric)\} is considered, 
where $R_{ax}/L_n\! =\! 2$, $R_{ax}/L_{T_{\mtr{i}}}\! =\! 8$, and $T_{\mtr{i}}/T_{\mtr{e}}=1$. 
In the simulations, the geodesic curvature $\mathcal{K}_\mtr{g}$ is artificially modulated in order to extract the geometric dependence, 
and the other geometrical quantities are fixed. 
%

%
Figure 1 shows the residual zonal-flow(ZF) level 
$R_{k_{\mtr{zf}}} := \left < \dphi_{k_{\mtr{zf}}}(t) \right >_z / \left < \dphi_{k_{\mtr{zf}}}(t=0) \right >_z$ 
as a function of the geodesic curvature $\mathcal{K}_{\mtr{g}}$ 
evaluated by the linear gyrokinetic calculations for the zonal component at $t=50R_{\mtr{ax}}/\vti$, 
where the quantities are normalized by the reference values corresponding to the original magnetic configuration 
without any modulations to $\mathcal{K}_{\mtr{g}}$. 
Such normalization enable us to compare the various magnetic configurations in the context of relative amplification and attenuation. 
One finds that the similar amplification/attenuation behavior of the residual zonal-flow level appears in all configurations considered here, 
where a power-decay function $X^{-\alpha}$ with the exponent of $\alpha=1.5$ well approximates $\mathcal{K}_{\mtr{g}}$ dependence.  
%
%
\begin{figure}[t]
  \centering
  \includegraphics[scale=1.2]{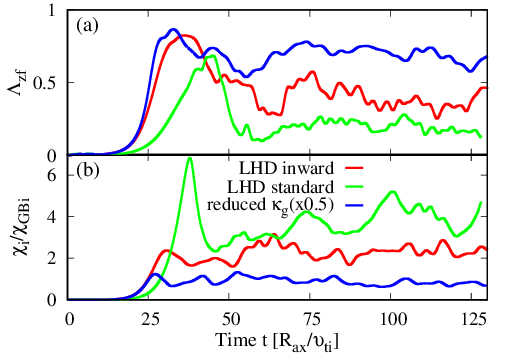}
  \caption{Time evolution of (a)relative zonal-flow intensity $\Lambda_{\mtr{zf}}$ and (b)turbulent transport diffusivity.}
  \label{fig:somefig}
\end{figure}
Then the nonlinear gyrokinetic simulations of the ITG-driven turbulence have been carried out to examine 
the zonal-flow amplification for the case with a reduced $\mathcal{K}_\mtr{g}$.
The LHD standard configuration, the LHD inward-shifted configuration, and its modulated configuration 
with a reduced $\mathcal{K}_\mtr{g}/\mathcal{K}_{\mtr{g}}^{\mtr{(ref)}}=0.5$ are considered as representative cases.   
As shown in Figs. 2(a), the relative zonal-flow intensity defined by $\Lambda_{\mtr{zf}}:= \mathcal{Z}/(\mathcal{T}+\mathcal{Z})$ strongly 
depends on the magnetic configuration, where $\mathcal{Z}:= \sum_{k_x} |\dphi_{kx,ky=0}|^2$ 
and $\mathcal{T}= \! \sum_{k_x}\! \sum_{k_y \neq 0}|\dphi_{kx,ky}|^2$. 
One can confirm the previous finding that the inward-shifted case shows stronger zonal-flow generation 
in comparison to the standard case\cite{thw_zf}, where the turbulent transport is reduced, as shown in Fig. 2(b). 
It is also emphasized that a more significant amplification of the relative zonal-flow intensity ($\Lambda_{\mtr{zf}} \! \sim\! 0.75$ 
) is observed in the case with reduced $\mathcal{K}_{\mtr{g}}$. 
Note that, as indicated for $t<25 R_{\mtr{ax}}/\vti$ in Figs. 2, the modulation of $\mathcal{K}_{\mtr{g}}$ has almost no impacts 
on the linear ITG growth because the operator $\mathcal{K}_\mtr{g}k_x$ in Eq. (2) does not act on the most unstable eigenmode with $k_{x}=0$. 
The zonal-flow amplification is, thus, attributed from the improved response $R_{k_{\mtr{zf}}}$ induced by the geodesic curvature 
$\mathcal{K}_{\mtr{g}}$. 

The results of systematic scans for $\Lambda_{\mtr{zf}}$ with respect to $\mathcal{K}_{\mtr{g}}$ are summarized in Fig.3, 
where the three kinds of configurations are used as the reference.     
Although we see a slightly different geodesic curvature dependence of the residual zonal-flow levels (see Fig. 1), 
a qualitatively similar power-like decay $\Lambda_{\mtr{zf}} \! \propto \! \mathcal{K}_{\mtr{g}}^{-\alpha}$ 
is identified even for quite different three magnetic geometry, where the exponent has some extent of $0.4 < \alpha < 1$. 
%

One possible reason for the difference of $\mathcal{K}_{\mtr{g}}$ dependence in $\Lambda_{\mtr{zf}}$ and $R_{k_{\mtr{zf}}}$ 
is that the zonal-flow intensity in the nonlinear state depends not only on the linear response, 
but also on the turbulence intensity(source of zonal flows) with various spectral shapes in the wavenumber space. 
Also, the geodesic acoustic modes(GAM) are not explicitly distinguished in this study, 
and such individual modeling for GAM and the steady zonal flows becomes important depending on the q-value. 
Further modeling studies including them are remained as future work. 

\begin{figure}[b]
  \centering
  \includegraphics[scale=1.2]{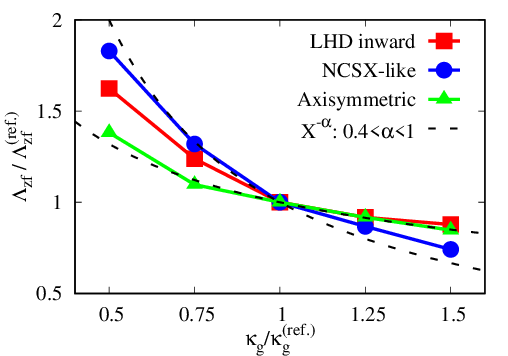}
  \caption{$\mathcal{K}_{\mtr{g}}$ dependence of relative zonal-flow intensity $\Lambda_{\mtr{zf}}$.}
  \label{fig:somefig}
\end{figure}
In conclusion, this paper has presented gyrokinetic simulations to extract 
the geodesic curvature dependence of nonlinearly generated zonal flows. 
The zonal-flow amplification by the magnetic geometry revealed here would be one of the plausible strategies towards a new discovery of 
innovative confinement geometry and the optimizations of magnetic configurations in the experiments. 
Indeed, utilizing the nonlinear turbulence modeling\cite{nunami} and the proxy approach for the linear ITG/TEM 
optimizations\cite{prox1,prox2,prox3}, we can derive a nonlinear proxy function with  
$\Lambda_{\mtr{zf}} \! \propto \! \mathcal{K}_{\mtr{g}}^{-\alpha}$,  
\begin{equation}
\chi_{\mtr{NL}}^{\mtr{proxy}}  = \frac{C_1 \left (\sum_{k_\perp}\gamma_{k_\perp}^{\mtr{proxy}}/k_{\perp}^{2} \right )^{\beta}}{1+ C_{2}\mathcal{K}_{\mtr{g}}^{-\alpha}/ \left (\sum_{k_\perp}\gamma_{k_\perp}^{\mtr{proxy}}/k_{\perp}^{2} \right )^{\delta}}, 
\end{equation}
where ($C_1, C_2, \alpha, \beta, \delta$) are optimal parameters to be determined\cite{nakayama}. 
The nonlinear proxy model encourages the explorations of novel magnetic geometry to activate zonal-flow dynamics, 
as will be reported in future work. 
%

%
The authors thank NGS(Next Generation Stellarator) creation research team at NIFS for their fruitful discussions. 
This work is supported by the MEXT Japan, 
Grant No. 19K03801, in part by the NIFS collaborative Research Programs(NIFS22KIST017, NIFS22KIST010), 
and in part by JST, PRESTO Grant Number JPMJPR21O7, 
and in part by PLADyS, JSPS Core-to-Core Program. 
Numerical simulations were performed by JFRS-1 at IFERC-CSC, and Plasma Simulator at NIFS.

%
%


\end{document}